\documentstyle[aps,epsf,eqsecnum]{revtex}
\begin{document}

\title{A soluble statistical model for nuclear fragmentation}

\author{S. Das Gupta$^{1,2}$, A. Majumder$^1$, S. Pratt$^2$ and A. Mekjian$^3$}

\address{ $^1$Physics
Department, McGill University, 3600 University St.,\\
Montr{\'e}al, Canada H3A 2T8\\
$^2$Department of Physics and National Superconducting Cyclotron Laboratory\\
Michigan State University, East Lansing, MI 48824\\
$^3$Department of Physics, 126 Frelinghuysen Road,\\
Rutgers University, Piscataway, NJ 08854}

\date{ \today}

\maketitle

\begin{abstract}
The statistical model of Chase and Mekjian, which offers an analytic solution
for the canonical ensemble of non-interacting fragments, is investigated for
it's thermodynamic behavior. Various properties of the model, which exhibits a
first-order phase transition, are studied. The effects of finite particle
number are investigated. Three extensions of the model are considered,
excluded volume, Coulomb effects and inclusion of isospin degrees of freedom. A
formulation of a microcanonical version of the model is also presented.

\end{abstract}


\section{INTRODUCTION}

Nuclear fragmentation resulting from heavy ion collsions is a complex
phenomenon. The role of equilibration and dynamics has not yet been determined
as a plethora of approaches have been investigated. Examples of approaches
are evaporative pictures\cite{friedman}, percolation models\cite{bauer,campi},
lattice gas models, and dynamical models based on Boltzmann
simulations\cite{chomaz,feldmeier,ohnishi,kiderlen,montoya}.  In this paper we
consider the statistical approach\cite{randrup,copenhagen} where one considers
sampling all configurations of non-interacting clusters. Recently, Chase and
Mekjian\cite{chase} derived relations which allow the exact calculation of the
canonical partition function for such a system.  By eliminating the need for
computationally intensive Monte Carlo procedures and associated approximations,
this technique allows a deeper insight into the thermodynamic principles which
drive the statistics of fragmentation.

In the next section we present the recursive technique of Chase and Mekjian and
review the thermodynamic properties, some of which have already been presented
in the literature. We emphasize that the surface energy is the most important
parameter in determining the fragmentation and phase transition properties of
the model. In the three subsequent sections, we present extensions of the model
which are necessary for serious modeling of nuclear systems: excluded volume,
Coulomb effects, and isospin degrees of freedom. In section \ref{micro_sec} we
show how a microcanonical distribution may be generated from the canonical
distribution.

\section{The Model}

For completeness, we present an outline of the model, which is based on the
work of Chase and Mekjian\cite{chase}.  The expressions used here are based on
a picture of non-interacting liquid drops. Mekjian and Lee had also 
applied similar recursion relations\cite{leemekjian} to a more algebraically
motivated fragmentation model that was not based on a liquid-drop picture.

We consider that there are $A$ nucleons which thermalize in a volume $V$ much
larger than $V_0$ where $V_0=A/\rho_0$ is the ground state volume of a nucleus
of $A$ nucleons.  These nucleons can appear as monomers but also as composites
of $a$ nucleons.  The canonical partition function of this system can be
written as
\begin{eqnarray}
\Omega_A=\sum_{\langle \Sigma n_ka_k=A\rangle}
\Pi_{k}\frac{\omega_k^{n_k}}{n_k!}
\end{eqnarray}
where $\omega_k$ is the partition function of a single composite of size $a_k$,
$n_k$ is the number of such composites and the sum goes over all the partitions
which satisfy $\sum n_ka_k=A$. A priori this appears to be a horrendously
complicated problem but $\Omega_A$ can be computed recursively via the formula,
\begin{equation}
\Omega_A=\frac{1}{A}\sum_k\omega_k\Omega_{A-a_k}
\end{equation}
Here $\Omega_0$ is 1.  It is this formula and the generalisation of this to
more realistic case (see later) that makes this model so readily soluble.

All properties of the system are determined by the partition functions of
indepedent particles. The recursive formula above allows a great deal of
freedom in the choice of partition functions for individual fragments,
$\omega_k$. Any function of temperature, density and $A$ is allowed. However,
explicit dependence on the configuration of the remainder of the system is
outside the scope of this treatment.

For the illustrative purposes of this section, we assume the form,
\begin{equation}
\omega_k=\frac{V}{\hbar^3}\left(\frac{a_kmT}{2\pi}\right)^{3/2}
\times e^{-F_{k,{\rm int}}/T}
\end{equation}
The first part is due to the kinetic motion of the center of mass of the
composite in the volume $V$ and the second part is due to the internal
structure. Following the choice of reference\cite{bondorf} we assume the
form
\begin{equation}
\label{bondorf_fe_eq}
F_{k,{\rm int}}=W_0a_k-S(T)a_k^{2/3}-T^2a_k/\epsilon_0
\end{equation}
Here $W_0$ is the volume energy per nucleon(=16 MeV), $S(T)$ is the surface
tension which is a function of the temperature $T$.  The origin of the
different terms in Eq. (\ref{bondorf_fe_eq}) is the following:
$-W_0k+Sk^{2/3}$ is the ground state energy of the composite of $k$ nucleons,
and the last term in the exponential arises because the composite can be not
only in the ground state but also in excited states which are included here in
the Fermi-gas approximation.  Following reference \cite {bondorf} the value of
$\epsilon_0$ is taken to be 16 MeV.  Lastly the temperature dependence of
$S(T)$ in ref\cite {bondorf} is $S(T)=S(0)[(T_c^2-T^2)/(T_C^2+T^2)]^{5/4}$ with
$S(0)=18$ MeV and $T_c=18$ MeV.  Any other dependence could be used including
a dependence on the average density.

\begin{figure}
\epsfxsize=0.75\textwidth \centerline{\epsfbox{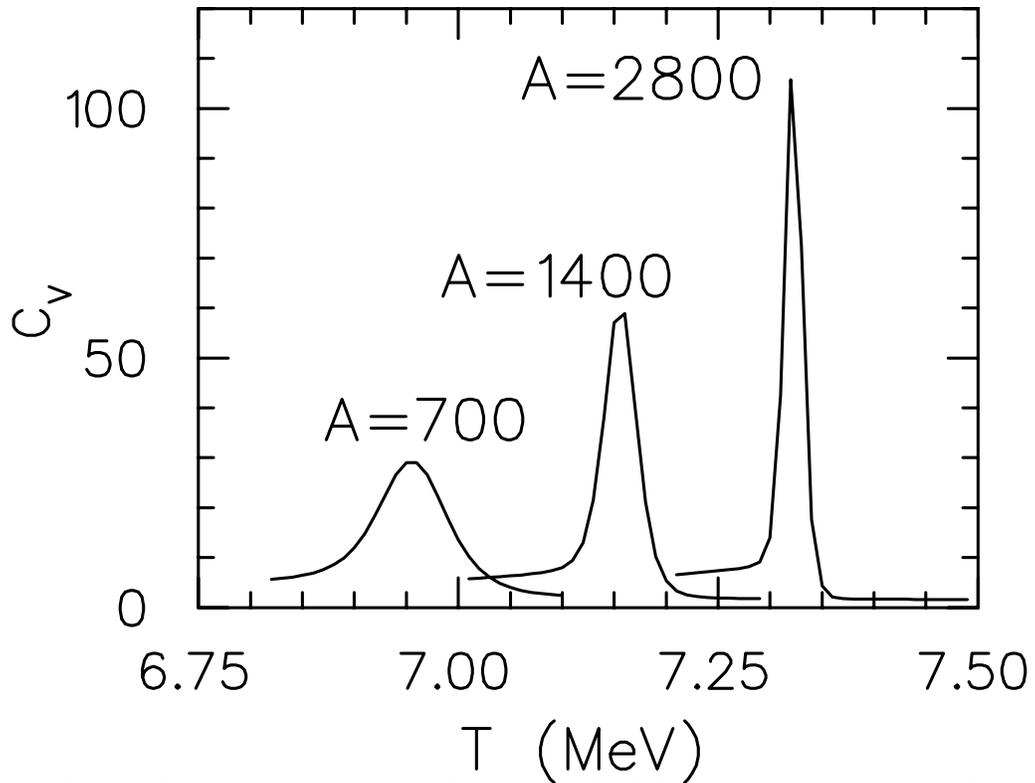}}
\caption{\label{cv_fig}
The specific heat is shown for three system sizes. As the size is
increased, the peak becomes increasingly singular. This demonstrates a
discontiuity in the energy as a function of temperature for a system held at 
fixed volume.}
\end{figure}

Upon calculation, the model described above reveals a first order phase
transition. In Figure \ref{cv_fig} the specific heat at constant volume,
$C_V=(1/A)dE/dT$, is displayed as a function of temperature for systems of
size, $A=700$, $A=1400$ and $A=2800$. The sharp peak represents a discontinuity
in the energy density, which sharpens for increasingly large systems. The usual
picture of a liquid-gas phase transition gives a discontinuity in the energy
density when pressure is kept constant rather than when the volume is kept
constant. To understand this result we consider a system divided into one large
cluster and many small clusters. The pressure and free energy may then be
approximated as
\begin{eqnarray}
E/A&\approx&\epsilon_{\rm bulk}+\frac{3}{2}\frac{N_{\rm cl.}}{A}T~,\\
\nonumber
P&=&\frac{N_{\rm cl.}}{V}T~,
\end{eqnarray}
where $N_{cl}$ is the number of clusters. The bulk term depends only on the
temperature and not on the way in which the nucleons are partioned into
fragments. We have neglected the surface energy term which is proportional to
$A^{-1/3}$. In this limit, $C_v$ and $C_p$ become
\begin{eqnarray}
C_V&=&\frac{\partial\epsilon_{\rm bulk}}{\partial T}
+\frac{3}{2}\frac{N_{\rm cl.}}{A}\\
\nonumber
C_p&=&C_V+\frac{N_{\rm cl.}}{A}.
\end{eqnarray}
The bulk term depends only on the temperature and is therefore continuous
across the phase transition. Thus, a spike in $C_p$ is equivalent to a spike in
$C_V$ since both are proportional to $N_{\rm cl.}$. It is difficult to make a
connection between this approach and the standard Maxwell construction, since
here interactions between particles enter only through the surface term.

\begin{figure} 
\epsfxsize=0.75\textwidth \centerline{\epsfbox{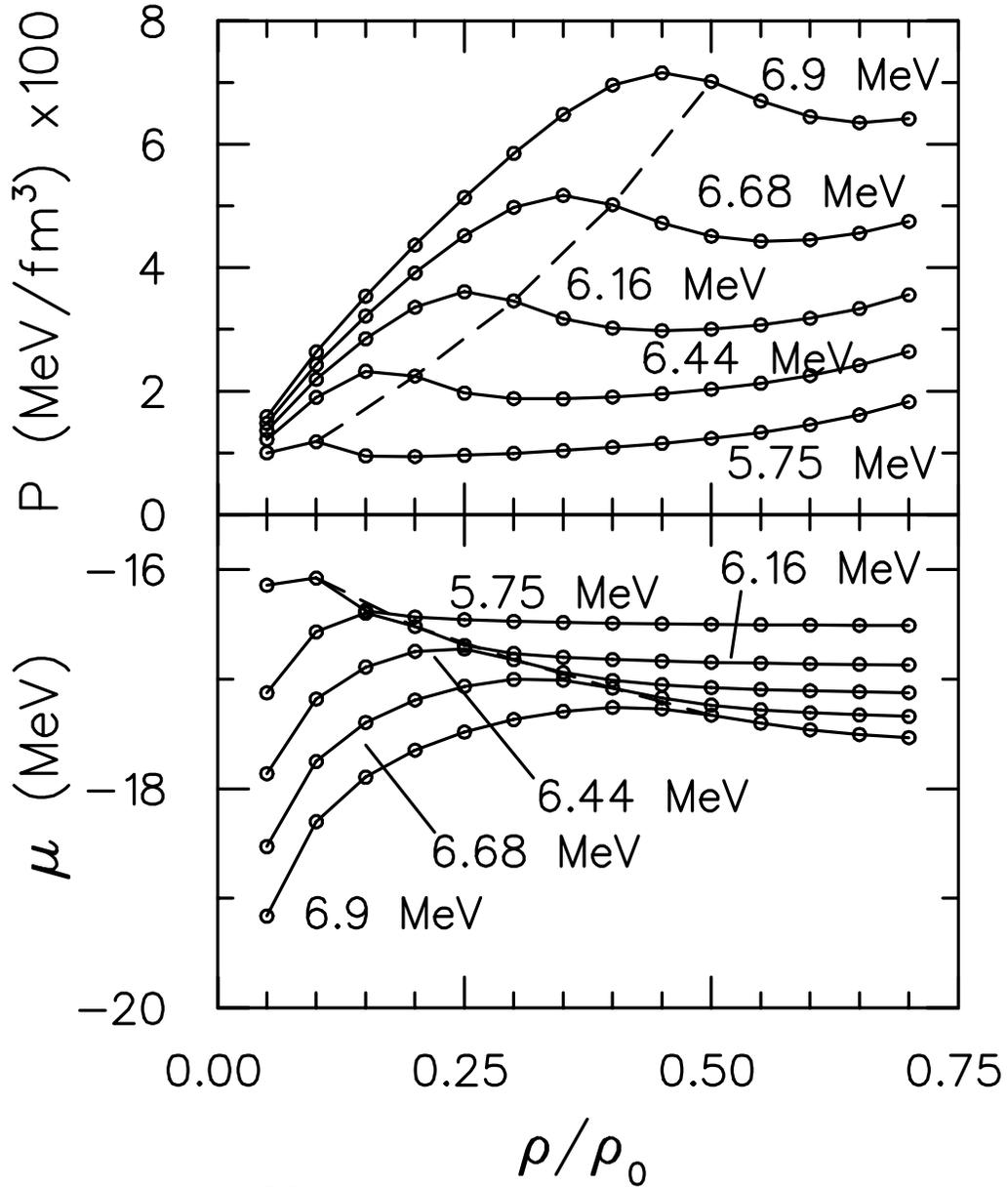}}
\caption{\label{mup_fig}
The $p-\rho $ diagram for $A$=200 at selected temperatures in the
upper panel shows the flattening of the pressure throughout a range of
densities, as expected in a first order phase transition. The dashed line shows
where $C_v$ is maximized as a function of temperature. The evolution of the
chemical potential is displayed in the lower panel.}
\end{figure}

Intrinsic thermodynamic quantities may be calculated in a straightforward 
manner. For instance the pressure and chemical potentials may be calculated 
through the relations,
\begin{eqnarray}
\mu&=&-T\left( \ln{\Omega_A}-\ln{\Omega_{A-1}}\right)\\
\nonumber
P&=&T\frac{\partial \ln(\Omega_A)}{\partial V}
\end{eqnarray}
Calculations of $\mu$ and $P$ are displayed in Figure \ref{mup_fig} as a
function of density for a system of size $A=200$. Both the pressure and
chemical potential remain roughly constant throughout the region of phase
coexistence. Of particular note is that the pressure actually falls in the
coexistence region due to finite size effects. 

We now make some comments about influences of various factors in
Eq. (\ref{bondorf_fe_eq}). The bulk terms, $W_0+T^2/\epsilon_0$, are not
affected by the free energy, thus they may be ignored when calculating
fragmentation observables. Their influence with respect to intrinsic
thermodynamic quantities is of a trivial character. The surface term $S(T)$ is
completely responsible for determining all observables related to fragmentation
and therefore all aspects of the phase transition. Aside from the system size
$A$, fragmentation is determined by two dimensionless parameters. The first is
the specific entropy, $(V/A)(mT/(2\pi\hbar^2))^{3/2}$ and the second is the
surface term $S(T)/T$.

\begin{figure} 
\epsfxsize=0.75\textwidth \centerline{\epsfbox{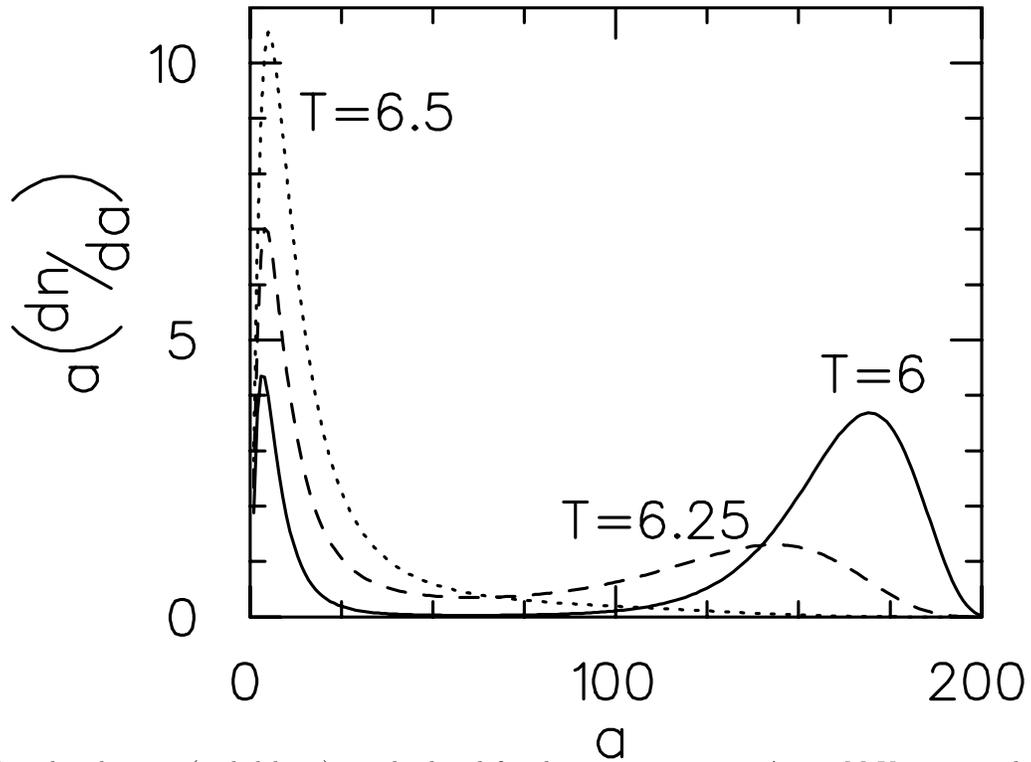}}
\caption{\label{massdist_fig}
Mass distributions (scaled by $a$) are displayed for three
temperatures. At 6.0 MeV, most nucleons reside in a single fragment, while at
6.5 MeV, most nucleons are part of small fragments. At the critical
temperature, 6.25 MeV, the mass distribution is remarkably broad.}
\end{figure}

At a given temperature the free energy $F=E-TS$ of $A$ nucleons should be
minimized.  With the surface tension term, $E$ is minmised if the whole system
appears as one composite of $A$ nucleons but the entropy term encourages break
up into clusters. At low temperatures the surface term dominates while at high
temperatures entropy prevails and the system breaks into small clusters. The
mass distribution may be calculated given the partition function.
\begin{equation}
\langle n_k\rangle=\frac{\omega_k\Omega_{A-a_k}}{\Omega_A}
\end{equation}
The mass distribution is displayed in Figure \ref{massdist_fig} for three
temperatures, 6.0, 6.25 and 6.5 MeV which are centered about the transition
temperature of 6.25 MeV. The distributions have been multiplied by $a_k$ to
emphasize the decomposition of the system. The mass distribution changes
dramatically in this small temperature range. The behavior is reminiscent of
that seen in the percolation or lattice gas models\cite{pan1}.

\section{Excluded Volume}

The volume used to define to the partition functions of individual fragments,
$\omega_k$ given in Eq. (\ref{bondorf_fe_eq}), should reflect only that volume
in which the fragments are free to move. Hahn and St\"ocker suggested using
$V\rightarrow V-A/\rho_0$ to incorporate the volume taken up by the nuclei. By
inspecting Eq. (\ref{bondorf_fe_eq}) on can see that this affects the partion
function by simply mapping the density or volume used to plot observables. More
realistically, the excluded volume could depend upon the multiplicity.
Nonetheless, in rather complicated calculations not reported here, it was found
that for the purpose of obtaining $p-V$ diagrams in the domain of interest in
this paper, it is an acceptable approximation to ignore the multiplicity
dependence of the excluded volume\cite {majumder}.

Incorporating a multiplicity dependence would be outside the scope of the
present model, as it would represent an explicit interaction between
fragments. However, one could add an $a$-dependence to the volume term to
account for the difficulty of fitting fragments of various sizes into a tight
volume. This might affect the model in a non-trivial fashion.

We like to remind the reader that the parameter $b$ in the Van der Waals EOS:
$(p+a/V^2)(V-b)=RT$ also has its roots in the excluded volume.  But there $b$
plays a crucial role.  We could not for example set $b$=0 without creating an
instability at high density. Furthermore, the phase transition disappears when
$a$ is set to zero.

\section{Coulomb effects}

It has been understood that the Coulomb effects alter the phase structure of
nuclear matter\cite{bali}. Although explicit Coulomb interactions are outside
the scope of this treatment, they may be approximated by considering a screened
liquid drop formula for the Coulomb energy as has been used by Bondorf and
Donangelo\cite{bondorf}. The addition to the internal free energy given in
Eq. (\ref{bondorf_fe_eq}) is
\begin{equation}
F_{\rm coul}=0.70\left(1-\left(\frac{\rho}{\rho_0}\right)^{1/3}\right)\frac{a_k^{5/3}}{4}~{\rm MeV}.
\end{equation}
This form implies a jellium of uniform density that cancels the nucleons
positive charge when averaged over a large volume. This may be more physically
motivated for the modeling of stellar interiors where the electrons play the
role of the jellium. 

\begin{figure} 
\epsfxsize=0.75\textwidth \centerline{\epsfbox{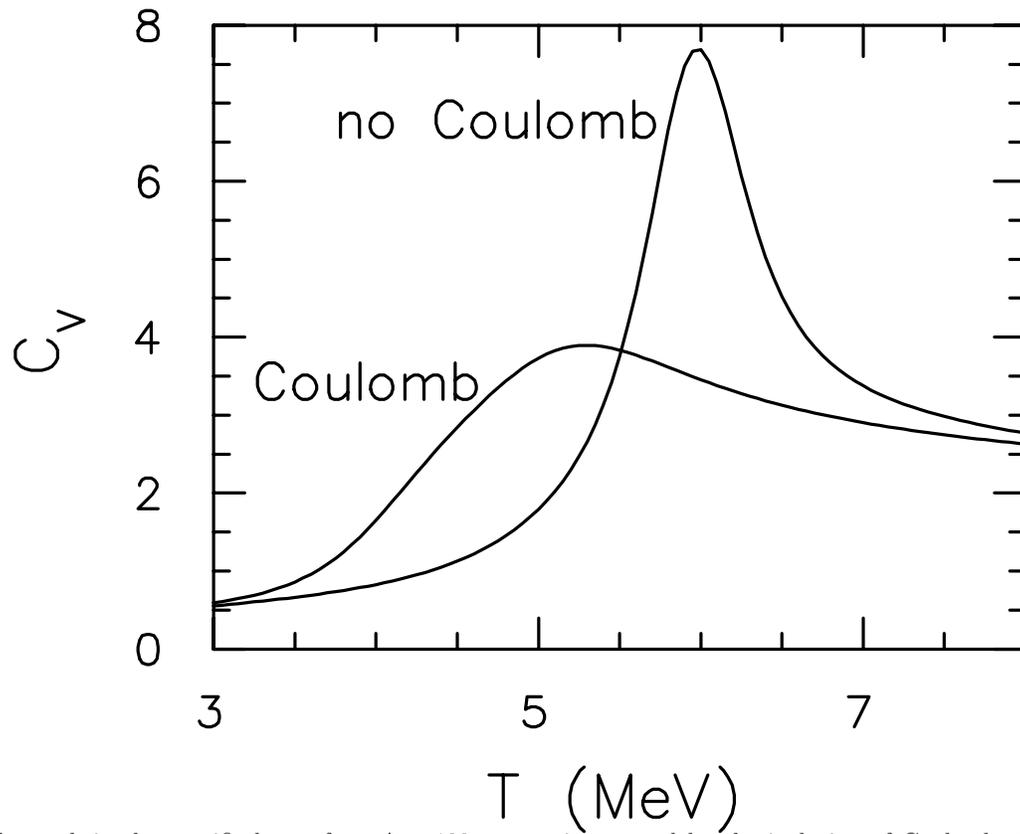}}
\caption{\label{cvcoulomb_fig}
The peak in the specific heat of an $A=100$ system is smeared by the
inclusion of Coulomb effects. For large systems, Coulomb destroys the phase
transition by making large drops energetically unfavorable.}
\end{figure}

We display $C_v$, both with and without Coulomb terms for an $A=100$ system in
Figure \ref{cvcoulomb_fig}. Coulomb forces clearly reduce the temperature at
which the transition occurs. For sufficiently large systems, Coulomb destroys
the transition as large drops become unstable to the Coulomb force.

\section{Conservation of Isospin}

The recursive approach employed here is easily generalized to incorporate
multiple species of particles. If there exist a variety of particles with
conserved charges $Q_1$, $Q_2\cdots$, one can write a recursion relation for
each charge\cite{chase_isospin}.
\begin{equation}
\Omega_{Q_1,Q_2\cdots}=\sum_k \frac{q_{i,k}}{Q_i}\omega(k)\Omega_{Q_1-q_{1,k},
\cdots Q_i-q_{i,k} \cdots}~,
\end{equation}
where $Q_i$ is the net conserved charge of type $i$ and $q_{i,k}$ is the charge
of type $i$ carried by the fragment noted by $k$.

For the nuclear physics example, one would wish to calculate $\Omega_{N,Z}$
where $N$ and $Z$ were the conserved neutron and proton numbers. To find
$\Omega_{Z,N}$ one must know $\Omega_{N^\prime,Z^\prime}$ for all $N^\prime<N$
or $Z^\prime<Z$. To accomplish this one must use both recursion relations.

\section{Obtaining the microcanonical distribution}
\label{micro_sec}

In nuclear collisions, one does not have access to a heat bath, but one can
vary the excitation energy. A microcanonical treatment is therefore more
relevant for practical calculations, particularly given the existence of a
first order phase transition which occupies an infinitesimal (in the limit of
large $A$) range of temperatures in a canonical calculation, but a finite
range of energies in a microcanonical ensemble.

The relevant partition function for a microcanonical ensemble is the density of
states, 
\begin{eqnarray}
\rho(E)&=&\sum_i \delta(E_i-E)\\
\nonumber
&=&\frac{1}{2\pi}\int d\beta 
\sum_i e^{i\beta (E-E_i)}\\ 
\nonumber
&=&\frac{1}{2\pi}\int d\beta ~\Omega(i\beta)e^{i\beta E},
\end{eqnarray}
where the sum over $i$ represents the sum over all many-body states.  Although
$\Omega(i\beta)$ is easily calculable given the recursion relations discussed
in the previous sections, one must perform the integral over $\beta$
numerically.

The true solution for the density of states would be ill-defined given the
discreet nature of quantum spectra which can not be combined with a delta
function. However, if one defines the density of states in a finite region of
size $\eta$, the density of states becomes well-behaved even for discreet
spectra. For that reason we more pragmatically define the density of states as
\begin{eqnarray}
\rho_\eta (E)&\equiv&\sum_i \frac{1}{\sqrt{2\pi}\eta}
\exp{-\frac{(E-E_i)^2}{2\eta^2}}\\
\nonumber
&=&\frac{1}{2\pi}\int d\beta ~\Omega(i\beta)
e^{i\beta E -\eta^2\beta^2/2}
\end{eqnarray}
One might have considered replacing the delta function by a Lorentzian rather
than by a Gaussian, but this would be dangerous given that the density of
states usually rises exponentially for a many-body system. The finite range
$\eta$ used to sample the density of states might correspond to the range of
excitation energies sampled in an experimental binning. In the limit
$\eta\rightarrow 0$, $\rho_\eta$ approaches the density of states.

As an example of a quantity one may wish to calculate with a microcanonical
approach, we consider the average multiplicity of a fragment of type $k$ in a
system whose total energy is within $\eta$ of $E$.
\begin{eqnarray}
\langle n_k\rangle_{\eta}(E)&=&\frac{ \sum_i n_{i,k}\frac{1}{\sqrt{2\pi}\eta}
\exp{-\frac{(E-E_i)^2}{2\eta^2}} }{\rho_\eta(E)}\\
\nonumber
&=&\frac{ \frac{1}{2\pi}\int d\beta \omega_k(i\beta)\Omega_{A-a_k}(i\beta)
e^{i\beta E-\eta^2\beta^2/2} }{\rho_\eta(E)},
\end{eqnarray}
where $n_{i,k}$ is the number of particles of species $k$ within the fragment
$i$. 

The integration over $\beta$ clearly provides an added numerical challenge that
increases for small $\eta$. For the purposes of generating a mass distribution,
one must perform this integration for every species. It might be worthwile to
consider estimating the integrals over $\beta$ with the saddle point method,
although one should be wary of taking derivatives of $\Omega$ with respect to
$\beta$ in the phase transition region.

Microcanonical quantities might also be calculated in a completely different
manner by discreetizing the energy. For instance one might measure energies in
units of 0.1 MeV. One might then treat energy on the same footing as any other
conserved charge. One may then write recursion relations for $N_{A,E}$, the
number of ways to arrange $A$ nucleons with net energy $E$, where $E$ is an
integer. 
\begin{eqnarray}
N_{A,E}&=&\sum_{k,e_k} \frac{a_k}{A}\omega_{k,e_k} N_{A-a_k,E-e_k}\\
&=&\sum_{k,e_k} \frac{e_k}{E}\omega_{k,e_k} N_{A-a_k,E-e_k}
\end{eqnarray}
Here, $\omega_{k,e_k}$ is the number of ways of arranging a fragment of type
$k$ with net energy $e_k$. All other relavant microcanonical quantities may be
calculated in a similar manner.

Since one needs to calculate $N$ at all energies $E^\prime$ less than the
targeted energy $E$, and must sum over all energies less than $E^\prime$ to
obtain $N_{A^\prime,E^\prime}$, the length of the calculation is proportional
to $E^2$. Typically, nuclear decays occur with on the order of a GeV of energy
deposited in a nucleus. Therefore, these calculations may become numerically
cumbersome unless the energy is discreetized rather coarsely.

\section{summary}

The recursive techniques discussed here have several attractive features.  They
are easy to work with, incorporate characteristics of nuclear composites and
appear to have standard features of liquid-gas phase transitions.  In the
present forms these models are resricted to low densities.  For modeling
nuclear disintegration this is not a serious problem, although for completeness
it would be nice to be able to modify the model so that it can be extended to
higher density. 

In this paper we have studied thermal properties of the model, and we emphasize
the importance of the surface term in determining these properties. We can
associate the discontinuity in the energy density with temperature to the
discontinuity in the number of clusters.  In addition, we have seen that
including Coulomb effects lowers the temperature at which the fragmentation
transition occurs and reduces the sharpness of the phase transition. We have
also presented an extension of the formalism for the inclusion of isospin
degrees of freedom.

For comparing to nuclear physics experiments, development of the microcanonical
approaches presented here is of greatest importance. It remains to be seen
whether the microcanonical formalisms are tenable, as they have yet to be
implemented. 

\acknowledgements{This work is supported in part by the Natural Sciences and
Engineering Research Council of Canada and by {\it le fonds pour la Formation
de Chercheurs et l'aide \'a la Recherche du Qu\'ebec\/}, by the US Department
of Energy, Grant No. DE FG02-96ER 40987, and by the US National Science
Foundation, grant 96-05207.}

\newpage

\end{document}